\documentclass[conference]{IEEEtran}

\newif\ifincluded
\includedfalse

\usepackage{cite}
\usepackage{amsmath,amssymb,amsfonts}
\usepackage{algorithmic}
\usepackage{graphicx}
\usepackage{textcomp}
\usepackage{xcolor}
\usepackage{hyperref}
\usepackage[hyphenbreaks]{breakurl}    

\begin{document}

\title{Sidechains and interoperability\\}

\author{\IEEEauthorblockN{Sandra Johnson}
\IEEEauthorblockA{\textit{Protocol Engineering Group and } \\
\textit{Systems (PegaSys), ConsenSys}\\
Brisbane, Australia \\
sandra.johnson@consensys.net}
\and
\IEEEauthorblockN{Peter Robinson}
\IEEEauthorblockA{\textit{Protocol Engineering Group and } \\
\textit{Systems (PegaSys), ConsenSys}\\
Brisbane, Australia \\
peter.robinson@consensys.net}
\and
\IEEEauthorblockN{John Brainard}
\IEEEauthorblockA{\textit{Protocol Engineering Group and } \\
\textit{Systems (PegaSys), ConsenSys}\\
Boston, USA \\
john.brainard@consensys.net}
}

\maketitle

\begin{abstract}
There appears to be an insatiable desire for spawning new bespoke blockchains to harness the functionality provided by blockchain technologies, resulting in a constant stream of blockchain start-up companies entering the market with their own unique vision and mission. Some target a particular niche market such as supply chain and financial services, while others strive to differentiate themselves from the increasingly saturated market by offering new functionality. This dynamic and constantly changing blockchain ecosystem makes it very challenging to keep abreast of all the latest breakthroughs and research. It is evident that there is also a growing desire to collaborate with others developing blockchain solutions, which brings new impetus to blockchain interoperability research. We review the strategies that some key players in the blockchain ecosystem have implemented, or are proposing to develop, to satisfy this increasing demand for cross-chain communication and transactions between sidechains. Interoperability presents a complex and challenging stumbling block to the wider uptake of blockchain technology. We find that although there is a plethora of blockchains and interoperability implementations, or proposals, at a higher level of abstraction there is only a handful of approaches. However, the way they are implemented can differ quite substantially. We present a summary of the reviews we conducted in a table for ease of comparing and contrasting. \end{abstract}

\begin{IEEEkeywords}
blockchain, sidechain, cross-chain, communication, interoperability
\end{IEEEkeywords}

\section{Introduction}
In public blockchains, such as Ethereum MainNet, all transactions and participants are visible, and any node can join the network. By contrast, private permissioned blockchains keep the transactions and participants secret and restrict the network to only approved nodes. The term `sidechains' is often used to refer to such blockchains and frequently used in the context of private consortium blockchains. This paper reviews the many techniques used for communicating between sidechains.

We will use the term `sidechain', although this technology may also be referred to in the literature as `satellite chains', `child chains', or `sub-chains', albeit that they may represent slightly different concepts, the overarching concept is similar in as far as they typically aim at taking traffic off a public blockchain, which increases transaction throughput and decreases latency, and pinning back to the public blockchain periodically to take advantage of its strong repudiation properties \cite{Li2017b}. The paper `Enabling Blockchain innovations with Pegged Sidechains' \cite{Back2014}, is generally perceived to be the first paper to be published about sidechains \cite{Lee2018}.

This review is not purporting to be all encompassing, but rather a snapshot at the time of writing of the key players that the authors are aware of, and of the technologies most closely related to the \textit{Ethereum private sidechains project} that the authors are involved in.

\section{Sidechains and Interoperability}

In reviewing the state of play we assimilated information from various sources: self-publication, technical reports, online articles, recorded videos of presentations, Github documentation, conference papers and academic journals. We summarise the main features of \textit{Ethereum Private Sidechains}, \textit{Plasma}, \textit{Polkadot} and \textit{Ethereum 2.0 Sharding}, as well as proposals and implementations for handling cross-chain interoperability. Additionally, we provide a brief overview of other interesting and relevant projects and implementations. We acknowledge that we may have overlooked some notable advances in this field, and endeavour to continually update our knowledge base as new information comes to light.

\subsection{Ethereum Private Sidechains}
The \textit{Sidechains} project at ConsenSys (PegaSys) is currently developing a Proof of Concept (PoC) to deliver on-demand, ephemeral, private, permissioned Ethereum sidechains that promises to ensure confidentiality of both in-flight and at-rest data. 

There are several novel concepts being introduced by this technology, such as being able to establish a sidechain between participants `on-demand', i.e. creating a unique sidechain in a matter of seconds or minutes, rather than days \cite{Robinson2018b}. The sidechain is created using domain name information which resides within the \textit{Ethereum Registry Authority} \cite{Robinson2018b}. Other authorised participants may join the sidechain once it has been created. Alternatively, participants may opt to leave the sidechain voluntarily, or they may be forced to leave if they are voted out by the other sidechain participants due to misbehaviour. 

Ethereum Private Sidechains leverage the non-repudiation properties of Ethereum MainNet, by periodically pinning the block hash of the sidechain state back to the public blockchain to prevent the sidechain participants colluding to revert the state of the sidechain \cite{Robinson2019}. The pinning approach gives participants the ability to contest posted pins without revealing the identities of all of the sidechain participants \cite{Robinson2019}. The system also allows for cross-chain transactions, allowing function calls between sidechains, atomically updating state across chains. Another novel concept is the ephemeral nature of an Ethereum Private Sidechain. The sidechain may only be required for a certain time, and thereafter no activity will take place on the sidechain. In this situation, a request is sent to archive the sidechain, which has to be verified by all the sidechain participants, before pinning the final state back to Ethereum MainNet and archiving other sidechain specific information. All of the features of Ethereum Private Sidechains have been designed to restrict the list of participants of the chain using permissioning, keeping the identities of the participants secret, and upholding the decentralisation ethos of the Ethereum platform. 
 
\subsection{Plasma}

Buterin and Poon published a white paper to propose the Plasma framework using Proof of Stake (PoS) \cite{King2012} with economic incentives for chain participants to exhibit good and efficient behaviour, and to operate autonomously \cite{Poon2017}. The motivation for Plasma is primarily scalability and runs `child chains' which reports back to the Ethereum MainNet periodically \cite{Zhang2018}. It minimises the data required to confirm transactions, organising Plasma chains into a tree hierarchy and using MapReduce functionality to speed up computation \cite{Poon2017}. Plasma chains can be private or public blockchains \cite{Poon2017}. 

The key characteristics of a Plasma framework are: smart contracts with cryptographic verification, transactions running on child chain instead of MainNet, periodic reporting back to Ethereum MainNet, dispute settlement uses mathematically verifiable methods to reward the correct party, and ensuring that only the owner can retrieve funds.

The first implementation of Plasma, Minimal Viable Plasma (MVP), is a simple unspent transaction output (UTXO) chain with the majority of the work being done `off-chain', i.e. on the sidechain rather than on Ethereum MainNet  \cite{Buterin2018}. MVP blockchains use proof-of-authority (PoA) for consensus. The smart contract only deals with deposits and withdrawals which are the entry and exit points, with all other transactions being performed on the sidechain  \cite{LearnPlasma}. OmiseGO announced plans to use Plasma MVP to build a decentralised exchange (DEX), and mentioned that only PoA consensus was available at that stage, but that PoS would be added later \cite{Knott2018}. Plasma MVP exit criteria requires immediate exit from a child chain if an invalid transaction has been included. Moreover, the child chain relies on Ethereum for block finality \cite{Knott2018}.

Another version of a Plasma implementation is Plasma Cash. Using the word `cash' reflects the concept of cash denominations, for example \$100, \$50 or \$20 bank notes, which all have a unique serial number \cite{LearnPlasma}. It has non-fungible tokens (NFTs), since when a user deposits Ether (ETH) into a smart contract they get a unique identifier so that they are able to store information pertaining only to their tokens, or coins, and this deposit cannot be merged with another deposit and cannot be divided into smaller denominations. The transactions are ``stored in either a simple sparse Merkle tree or a patricia tree, with the index being the ID of the coin that is spent'' \cite{Buterin2018a}. 

Loom built PlasmaChain (aka Zombie Chain), a hub that bridges multiple sidechains into Ethereum, to enable faster and cheaper token transactions with a built-in decentralised exchange (DEX), and a more performant chain for developers to deploy their Solidity DApps. In addition to PlasmaChain, Loom created GameChain for interactive mobile games and SocialChain for social applications \cite{Duffy2018}. Loom is currently running all three in production as sidechains, linked to Ethereum MainNet by Plasma Cash for ``native ETH, ERC20, and NFT (ERC721) support'' with plans to integrate bitcoin payments \cite{Duffy2018}. Moreover, they propose the concept of ``sidechains of sidechains'' where the role currently being performed by Ethereum MainNet will be replaced by a PlasmaChain parent chain so that transactions across the various sidechains are processed without having to pin back to MainNet \cite{Duffy2018}. This concept is analogous to Beacon chains in Ethereum 2.0 and the Relay Chain in Polkadot \cite{Wood2017,Edgington2018}. 

Loom also introduced the concept of a DAppChain, which is a DApp running as its own sidechain \cite{Konstantopoulos2018}. For example, DelegateCall.com is a DAppChain running a SocialChain sidechain on the Loom Network \cite{Duffy2018}.

The Plasma XT implementation provides the means to reduce storage requirements and to maintain crypto-economic aggregate signatures and checkpointing. A successful checkpoint enables a user to discard the history prior to this checkpoint \cite{Zhang2018, Fichter2018}. Other implementations are Plasma Debit and More Viable Plasma. Plasma Debit allows the exchange of divisible amounts, which were not possible in any of the previous Plasma frameworks.

Choice of the most suitable version of Plasma depends on each particular use case as there are many intricacies of the technology. Some key challenges remain, such as infinite scaling and mass exiting from a child chain under certain circumstances. Careful consideration of time to challenge, and operators fronting money also need to be taken into account.

\subsection{Polkadot}
The Polkadot network topology consists of \textit{relay chains}, \textit{parachains} and \textit{bridges}. A relay chain coordinates consensus and transaction delivery between chains, parachains receive and process transactions, and bridges are special parachains that link two chains with their own consensus \cite{Wood2017}. 

Parachains are added to the network by bonding tokens which can be removed when the parachain is no longer relevant \cite{Web3Foundation2017}.
Polkadot uses a Proof of Stake (PoS) interface contract by taking the Proof of Authority (PoA)  consensus mechanism and turning it  into a PoS mechanism with logic to decide who the decision-making authorities are based on criteria such as DOT tokens staked, when they were staked, and the reward taken for their work \cite{Wood2017a}. Polkadot smart contracts are coded in languages that compile to WebAssembly (Wasm), with Rust the preferred language \cite{Wood2017a}. 

The primary parachain protocol is a parachain that exists within the native Polkadot infrastructure and has a Wasm execution environment. Polkadot proposes two protocol alterations: the first is to have `chromatic gas' that would vary depending on requirements, e.g. computation or storage. The second protocol alteration is to remove `dust accounts', which are accounts with no real value attached to them \cite{Wood2017a}.

Bridging enables communication between Polkadot and legacy or proprietary blockchains that already have their own consensus mechanisms in place, such as Ethereum, via `break-in' and `break-out' contracts. These contracts (aka gateway contracts) enable Ethereum to forward and receive messages from the outside world into Ethereum smart contracts \cite{Wood2017a}. Bridging uses light client block proofs (essentially Ethereum light clients) to guarantee, in the absence of knowledge of the current blockchain state, that a transaction executed as expected, or that ``a particular part of the state has a particular piece of data in it'' \cite{Wood2017a}.  Polkadot bridging uses a multi-signature system to enable validators to operate as a whole, coordinating and cooperating to spend jointly owned funds. Validators may own funds in another network, e.g. Ether in Ethereum, or BTC in Bitcoin. Each of the validators, or a subset of validators, has part of a key which is used to coordinate the signing of a transaction on behalf of the full set of validators. In order to spend validator funds a predetermined fraction (e.g. $\tfrac{2}{3}$) of the full set of validators has to approve the transaction to spend the funds. Polkadot assumes that no more than 20\% of the validators are nefarious, but if there are more than 20\% bad validators, then the spend can be blocked. However, if 80\% or more are bad, then the funds can be spent. This decentralised secret store and signer is ``effectively a way of allowing authority or validators nodes to be able to own an address without any one of them owning that address.'' \cite{Wood2017a}  

The Polkadot network is maintained by two key classes of participants: \textit{collators} and \textit{validators} with complementary roles being performed by \textit{nominators} and \textit{fishermen} \cite{Web3Foundation2017}. 

Each collator manages a specific parachain, and performs functions analogous to miners. They fish around for parachain transactions, bundle them together, and provide a proof of validity so that a validator (which doesn't maintain any particular parachain), regardless of the state they are in, can check that the block is valid. The collators in all the parachains provide proposals for blocks, so that there may be many blocks coming from each of the parachains, but only one block from each of the parachains must to be chosen to be validated and put into the finalised block. This is the task performed by a `parallelised and decentralised multi-chain aggregator' to provide the means by which validators can agree on which block from each of the parachains will be included in the final block of the relay chain \cite{Wood2017a}.

Collators submit state transition proofs to the validators. Similar to fishermen, they monitor the network for misbehaviour that they then prove to the validators. Validators route cross-chain messages by providing proofs that a particular parachain has a particular output queue, and proofs that a transaction in that output queue should be in another parachain's input queue, and in the next block. Validators secure the relay chain by staking DOT tokens. They validate proofs from the collators and participate in the consensus. Nominators stake DOTs when nominating validators to ensure the selection of honest validators. Fishermen primarily monitor the network and prove misbehaviour to validators \cite{Web3Foundation2017}.

In the relatively early stages developing Polkadot the decision was taken to split the codebase into two distinct repositories: Substrate and Polkadot \cite{ParityTechnologies2018d}. Substrate aims to be a generic framework for creating blockchains, providing the functionality to perform on-chain upgrades and to run a Light Client. Polkadot is a Rust Substrate implementation\cite{ParityTechnologies2018b}. Substrate uses libp2p for peer to peer networking and is adopting a progressive consensus methodology, leveraging the fact that every block in a blockchain references the entire history of the blocks before it. Therefore, a validator’s signature on one block also provides security on all the blocks preceding that block. The proposed Substrate consensus model is a hybrid consensus model of Aurand providing predictable chain growth with probabilistic finality, and practical Byzantine fault tolerance (PBFT) providing full, deterministic finality \cite{ParityTechnologies2018a}. Substrate has launched two test nets: BBQ Birch, which includes a Polkadot smart-contract runtime module and Charred Cherry, a testnet for Substrate 1.0 Beta which includes a combined Aura/GRANDPA consensus algorithm \cite{ParityTechnologies2018b}. 

Polkadot has launched three test nets: PoC-1 had a Wasm-based state-transition execution engine and dynamic on-chain upgrades, used to create the PoC-2 test net, Krumme Lanke. The PoC-3 test net, Alexander, includes the GRANDPA (GHOST-based Recursive Ancestor Deriving Prefix Agreement) finalisation algorithm, improved cryptography and governance logic such as lock-voting and delayed-enactment. GRANDPA provides instant finality under favourable conditions and eventual finality under adverse conditions \cite{Wood2018}. The fourth test net, due for release in April 2019, should  include inter-chain message passing (ICMP) \cite{Wood2018}.

\section{Sharding - Ethereum 2.0}
The ability to scale is an important goal of Ethereum 2.0, and the proposed way to achieve scaling is through sharding. Adapting the concept of database sharding, and borrowing from the Omniledger sharding design \cite{Kokoris-Kogias2018}, the sharding proposal for Ethereum is to split the network's computational requirements into shards, each with a capacity as high as the current Ethereum 1.0 chain. Every node in the Ethereum 2.0 network will not have to ``process (download, compute, store and read) every transaction in the history of the blockchain in order to make (write and upload) a new transaction'' \cite{Edgington2018b,EthereumFoundation2018}. 

Ethereum shard chains can be viewed as a special type of sidechain. Integral to the functioning of Ethereum shard chains is the PoS beacon chain \cite{Edgington2018}. The beacon chain maintains a registry of validators and their deposits, rewarding well behaved validators, but removing or penalising misbehaving validators \cite{Edgington2018,Ethereum.org2018}. It generates random numbers based on a RANDAO scheme to allocate block proposers and to assign validators to shards from the pool of validators \cite{Edgington2018, Ethereum.org2018,Skidanov2018}. The two key roles that validators perform are as a block proposer and as an attester \cite{Ethereum.org2018}. 

Network participants can become validators in Ethereum 2.0 once a deposit receipt of 32 ETH into an Ethereum 1.0 deposit contract is processed by the beacon chain \cite{Ethereum.org2018}. Amounts much larger than 32 ETH may be staked to acquire more seats in the validator pool. For example, if a participant stakes 320 ETH then 10 validator entities will be added to the pool. However, since validators are assigned randomly to shard chains, an attacker would need to put up a high proportion of the total stake in order to manipulate a shard chain by having a majority of the validators operating on that chain \cite{Skidanov2018}. Moreover, functionality will be added to the beacon chain to appoint shard sub-committees at random, whose task it is to ensure that validators are behaving correctly \cite{Edgington2018}. Apart from the requirement that Ethereum 2.0 validators need to stake ETH in Ethereum 1.0, the beacon chain is not reliant on Ethereum 1.0 \cite{Edgington2018}

A validator is assigned to propose a block for a specific time slot (each slot is currently 6 seconds), and all the other validators in the shard's committee need to cast their vote, i.e. ``attest'' to it, by signing it \cite{Ethereum.org2018,Skidanov2018}. Rather than including the signature of every validator that attested to a block, an aggregate signature is included in the block by means of the BLS signature aggregation scheme (using 128-bit security BLS12-381 curve) for each epoch \cite{Ethereum.org2018,Skidanov2018,Drake2018}. At the same time as creating the aggregated signature, a crosslink is created between the shard chain and the beacon chain to include information about which validators voted for that shard block, their aggregated signature, and other data \cite{Drake2018}. To verify a crosslink poses very little overhead on the beacon chain, as the verification of any one crosslink is very quick (in the order of milliseconds). Moreover crosslinks are expected to be created on the beacon chain for every epoch (64 slots) and therefore confirms the shard chain up to that validated shard block. In this way a beacon chain keeps track of the updated state of the shard chains \cite{Ethereum.org2018,Drake2018}. ``Crosslinks also serve as infrastructure for asynchronous cross-shard communication'' \cite{Ethereum.org2018}.

Staking to the main chain from the beacon chain and shard chains will use the full Casper Friendly Finality Gadget (FFG) logic \cite{EthereumFoundation2018,Buterin2017}, and the fork choice rule for all chains will use the Latest Message Driven (LMD) Greediest Heaviest Observed Sub-Tree (GHOST) rule \cite{Ethereum.org2018, Tsao2018}.  LMD GHOST ``is a block-vote weighted function'' \cite{Tsao2018} and is reliant on the block that is referenced in the most recently accepted crosslink \cite{Ethereum.org2018}, ``treating those latest messages as votes to cast decisions on a potential beacon chain head'' \cite{Tsao2018}.

\section{Other Proposals}

\subsection{Blockchain Router Proposal}
Wang et al (2017) proposes a blockchain router to manage cross chain communication \cite{Wang:2017:BRC:3070617.3070634}. All communication between chains are routed via the blockchain router and the sub-chains connect to it using a cross-chain communication protocol. All participants in the blockchain router perform a specific role: validators `verify, concatenate and forward blocks', nominators contribute funds to validators they trust, surveillants monitor blockchain router activity and report bad behaviour, connectors facilitate the flow of information between sub-chains and the router: sending and receiving information, executing transactions, signing results and collecting them into blocks to send to the validators. Each connector maintains a full node for a specific sub-chain. Their delegated stake-PBFT (DS-PBFT) consensus algorithm is a variation of the PBFT algorithm to include validators' voting rights \cite{Wang:2017:BRC:3070617.3070634}.

To incentivise participation in the system The value of a participant's stake is subject to inflation, so the expectation is that they will participate to prevent their stake devaluing.  The rewards for participants in the blockchain router system varies according to the role that they play: validators get the largest reward, followed by collectors and then nominators \cite{Wang:2017:BRC:3070617.3070634}. Surveillants may also receive rewards, although they do not directly participate in the system, when they witness and report bad behaviour \cite{Wang:2017:BRC:3070617.3070634}.

\subsection{Clearmatics}
The Clearmatics research project, Ion, focuses mainly on interoperability between private permissioned blockchains \cite{Clearmatics2018}. The first release of Ion, Ion Stage 1, implemented cross-chain atomic swaps for ERC-223 tokens between two Ethereum chains which identified several practical issues with atomic swaps \cite{Clearmatics2018a}, including the need for a centralised exchange which is ideologically opposed to blockchain decentralisation. They explored ways in which the reliance on a third party, such as an exchange, can be removed and tokens swapped directly between the two chains. The various scenarios they examined all presented challenges when one or both of the parties involved in the swap were nefarious. They presented use cases for a one-way swap, a bonded lock chain exchange, and a refunded hash-lock scenario to demonstrate the challenges for atomic swaps with malicious actors, and to ensure that the transaction is `rolled back' for incomplete or fraudulent swaps, and parties are refunded as required \cite{Clearmatics2018a}.

The second iteration of Ion provided a library of tools to enable the development of cross-chain smart contracts for blockchain interoperability, such as atomic swaps or decentralised exchanges, using the concept of `continuous execution' \cite{Clearmatics2018c}. The three key parts of Ion Stage 2 are a modular validation contract, an Ion hub contract and continuous execution contracts. 

Interoperability between two chains requires that the state is passed in a trustless manner between them. Within the Ion framework a cross-chain smart contract should only execute if it can be verified (via a smart contract) that a particular state transition has occurred. Therefore Ion provides interfaces for developers to create proofs of state for state validation and state transition verification \cite{Clearmatics2018c}. The state validation layer ensures that the blockchain state that has been passed has been checked for validity and correctness. Once the state has been passed, the state transition layer invokes code execution based on whether certain conditions have been met, e.g. account balance checking on another chain, ``or asserting that some transaction has been fulfilled'' \cite{Clearmatics2018c}. The concept of executing a function on one blockchain if and only if a specific state transition has been proved to have occurred on another blockchain, is referred to by Clearmatics as `continuous execution' \cite{Clearmatics2018}. Therefore, to verify that an expected computation took place, Ion checks for the presence of an event in a transaction and any further processing is dependent on the successful verification thereof \cite{Clearmatics2018c}. The Ion GitHub repository includes example functional cross-chain smart contracts for interoperability between Ethereum and Rinkeby, which uses a Clique PoA consensus for validation on Rinkeby Testnet \cite{Clearmatics2018c}.

\subsection{Metronome}
Metronome operates on Ethereum and is being developed by Bloq, whose co-founder, Jeff Garzik, is one of the early developers of Bitcoin \cite{Dale2018}. Metronome has a cross-chain currency token, MET \cite{Metronome2018, Dale2018}.  Metronome claims that it is:  (i) the first cryptocurrency that is not tied to a specific blockchain network (although as far as we are aware, it currently only runs on the Ethereum platform) and (ii) the first cryptocurrency that can potentially be secured by the ``best blockchain networks, without permanent commitment to any one blockchain'' \cite{Dale2018}. The owner’s manual \cite{Metronome2018} has a lot of detail on the processing and planned functionality of Metronome, but the one most relevant to this review is the portability aspect of Metronome across different chains \cite{Metronome2018}. The processing sequence for transferring Metronome from one blockchain, Blockchain A, to another blockchain, Blockchain B starts with the user initiating the transfer of MET tokens by committing to the target blockchain (Blockchain B), and obtaining a `proof of exit' merkle receipt from the smart contract when removing their tokens from the source chain (Blockchain A). The user then presents this receipt to the smart contract on the target blockchain (Blockchain B) to claim their MET tokens on Blockchain B \cite{Metronome2018}.  

The key components of Metronome are: exporting, importing, and validation \cite{Metronome2018}. The export function enables an owner of MET tokens to remove their tokens from the source chain and issues an export Merkle receipt. The owner pays the validator of this transaction a small fee. The import function on the destination chain processes the export receipt, which can be presented by any user, to deliver the MET tokens to the recipient on the target chain, and they can then claim the fee as witness of the transaction. Validation of import and export of MET tokens perform several duties: when there is a hard fork, validators need to attest to the valid chains, and they need to provide event proofs for import validation, amongst other responsibilities. Metronome has various phases of validation: phase 1 checks and validates export receipts; phase 2 creates a merkle tree of the hashes of the receipts, and validates the merkle root of the trees so that importers can provide a proof of import comprising of the ``merkle receipt and pairwise hashes attesting the root of the events'' to validators and users; phase 3 involves the validation of blockchain hashes of every chain that has MET tokens. Importers provide proofs by means of a merkle tree path to demonstrate that an export event is in a particular block header, and that the event in the block is the one that corresponds to the validated chain hashes \cite{Metronome2018}. It appears that the validation phases are still under development, with mention that a combination of Proof of Stake and Proof of Work are likely to be adopted, along with ``both soft and hard consensus for bad actors'' \cite{Metronome2018}.

\subsection{NEC Blockchain: Satellite Chains on Hyperledger Fabric}
Li et al. propose
a blockchain architecture where a `regulator' manages membership of sidechains (`satellite chains'), stores information about members (e.g. node ID, IP address, TCP port), contains a policy directory smart contract that enforces various policy smart contracts such as transaction validation and collection of validation results, and keeps a list of the sidechains registered to it \cite{Li2017b}. 

They claim to have successfully coded a PoC for sidechains which they integrated with the Hyperledger Fabric platform, running different consensus protocols in parallel with a regulator who oversees the entire network \cite{Li2017b}. The chains are essentially `sub-chains' of a main blockchain, with their own private ledger.

\cite{Li2017b}. A node will request to join a sidechain (satellite chain) if they want to transact with the other nodes on the sidechain. However, there is no suggestion on how this dynamic list of nodes attaching and detaching to the sidechain would operate. They claim to be able to run an `unbounded number of active chains', but again no in-depth explanation of how this is achieved, nor what is meant by `active' or `inactive' chains. 

Asset transfer between sidechains is via a direct connection between the sidechains using transport layer security (TLS) to send the transfer messages \cite{Li2017b}. The transaction payload includes the chain id of the target sidechain. The node receiving the transaction broadcasts the transaction to all the other participating nodes on the destination sidechain, and then invokes the relevant smart contract for the sidechain and transaction type \cite{Li2017b}.

The work presented in the paper \cite{Li2017b} appears to have been implemented by NEC according to the blockchain offerings on their website, including the transfer of assets between connected sidechains \cite{NEC2018}. They also claim to be able to achieve rates of around 100,000 transactions per second for a sidechain with 200 nodes. Moreover, the consensus is Byzantine Fault tolerant (BFT) using secure hardware, and sidechain participants are hidden from other sidechains, so that a stakeholder in one sidechain will be unaware of the existence of another sidechain and its participants \cite{NEC2018}. 

\subsection{Token Atomic Swap Technology (TAST) research project}
The Distributed Systems Group at TU Wien (Technische Universit\"{a}t Wien) in Vienna, Austria have written three white papers \cite{Borkowski2018d,Borkowski2018,Borkowski2018c} about their research into cross-chain communication. The first paper briefly reviews 20 blockchains that they consider to be the most prominent ones, and also 14 projects that are relevant to their proposed research project, TAST (Token Atomic Swap) \cite{Borkowski2018d}. They plan to introduce a cross-chain token, called PAN, which will be pivotal to their blockchain interoperability proposal. The challenges they identify that still need to be addressed by their research concern the issuance of tokens, the lifetime, states and balances of tokens, identification of suitable blockchains for their interoperability proposals and for cross-chain transfers, especially if there is a lack of ``Turing-complete smart contracts'' \cite{Borkowski2018d}. They identify the Metronome project \cite{Metronome2018} as being most closely aligned to their project and research, and the only other project in their list that uses a cross-chain token for asset transfers \cite{Borkowski2018d,Borkowski2018}. For the PoC they nominated Ethereum as one blockchain and the other blockchain as either Ethereum Classic, Bitcoin, Neo or Waves \cite{Borkowski2018d}. 

The second paper articulates the problems with a cross-blockchain proof and outlines a protocol for cross-blockchain asset transfers by means of `claim-first transactions' \cite{Borkowski2018}. They demonstrate in detail that it is not possible to verify on one blockchain, $C_A$, that data, $D$, have been recorded on another blockchain, $C_B$. They refer to this as the `cross-blockchain proof problem'. Stated formally in the paper as the `Lemma of Rooted Blockchains', the lemma requirements are that in order to verify from $C_A$ that specific data, $D$, are included in $C_B$ you need to be able to verify ``(i) the presence of a subset of the block lineage of $C_B$ on $C_A$, and (ii) the verifiability of the transaction consensus of $C_B$ by the transaction consensus of $C_A$'' \cite{Borkowski2018}. These are the typical problems with `spend first' transactions where the recipient of a transfer of funds, or assets, on one blockchain is unable to verify that the funds, or assets, have been marked as spent on the originating blockchain. Their proposal is to reverse this intuitive sequence of processing by creating a claim transaction for the funds on the target chain first. However, this introduces a different set of challenges \cite{Borkowski2018}:
\begin{itemize}
\item \textit{Proof of Intent (PoI)} - a proof to demonstrate the sender's intent to transfer the funds, which is a signed authorisation for the transfer of funds, using the sender's private key, which can then be verified publicly, at the time of posting the CLAIM transaction.
\item \textit{Balances} - to ensure that a SPEND transaction to transfer the funds or assets at the time that the CLAIM transaction was created
is possible, they propose that the account balances of all the participating parties, i.e. sender, receiver and witnesses, are recorded in all the blockchains that will be allowing cross-chain asset transfers. The SPEND transaction is created by a `witness' of the CLAIM transaction on the target chain.
\item \textit{Double spending} -  it is feasible that two CLAIM transactions could be created on two different blockchains for the same funds, if the spender signs more than one PoI, each on a different blockchain. To address double spending they suggest a finite time period for the validity of a PoI. Any overlapping PoIs from that sender will be marked as invalid by means of a VETO transaction.
\item \textit{Double destruction} - it is not infeasible that two witnesses of a CLAIM transaction on the target chain could create duplicate SPEND transactions on the source chain, which will mark as spent twice the amount of funds as should be spent. This situation could be circumvented by having a way of distinguishing between different instances of a PoI.
\item \textit{Reward Assets} - to incentivise witnesses to create SPEND transactions, they propose a witness reward, which could be paid from the transferred asset (e.g. a small fraction), or by having ``a dedicated reward asset''.
\end{itemize}

The third white paper presents a use case for a `claim-first' transaction, showing how `witness rewards' could be assigned, and providing an ``algorithm for creating a cryptographically verifiable PoI'', demonstrating that a cryptographic PoI for cross-chain asset transfer could be implemented \cite{Metronome2018}. Their initial PoC uses blockchains that have smart contracts, as they facilitate signature verification, which is needed for claim-first transactions because they are reliant on the PoI from the sender being verified \cite{Metronome2018}. Note that their claim-first transaction proposal requires balances of all the wallets to be present on all the participating blockchains, and not only on the target blockchain \cite{Metronome2018}. The assumption is that assets exist as tokens (e.g. ERC20 tokens) on the blockchains, independent of the blockchain’s native currency \cite{Metronome2018}. They state that only one non-malicious witness is necessary to maintain consistency across the blockchains \cite{Metronome2018}. When there is more than one witness of the claim-first transaction, the one with the shortest distance between the signature and the public key is the one who will win and be rewarded with a payment that uses the transferred asset \cite{Metronome2018}.   

\begin{table*}[htbp]
\caption{Summary of sidechain and cross-chain technologies reviewed}
\label{tab:summary}
\begin{center}
\begin{tabular}{|l|l|l|l|l|}
\hline
\textbf{\textit{Project}} & \textbf{\textit{Blockchain Features}}& \textbf{\textit{Interoperability}} & \textbf{\textit{Implementation status}} & \textbf{\textit{Platform}} \\
\hline
Blockchain router \cite{Wang:2017:BRC:3070617.3070634} &- participant roles: &- blockchain router &  no evidence of a PoC & unknown, \\
 &  \hspace{3pt} validators, nominators,  &\hspace{3pt} manages cross chain & & Bitcoin \& \\
 &  \hspace{3pt} surveillants, connectors & \hspace{3pt} communication via & \hspace{3pt} &Ethereum \\ 
 & - unclear whether private or   &\hspace{3pt} connectors linking router & & mentioned \\
 & \hspace{3pt} public & \hspace{3pt} to sidechains &   &  \\
 & - delegated  stake-PBFT & & &  \\
 & \hspace{3pt} consensus & & & \\
\hline
Clearmatics (Ion)  & - Clique PoA consensus & - atomic cross chain swaps & - PoC: cross-chain smart & Ethereum \\
 \cite{Clearmatics2018b,Clearmatics2018a,Clearmatics2018} & -  private and permissioned& \hspace{3pt} via DEX & \hspace{3pt} contracts between  & \\
 & \hspace{3pt}  blockchains &  - continuous execution & \hspace{3pt} Ethereum \& Rinkeby& \\
\hline
Ethereum private & - on-demand creation & - atomic function calls & - PoC in development & Ethereum \\
sidechains \cite{Robinson2018b, Robinson2019} & - private and permissioned & \hspace{3pt} for transactions and & & \\
 & - ephemeral with archiving & \hspace{3pt} messaging & & \\ 
 & - anonymous pinning to MainNet & - detailed specification& & \\
 & - Ethereum Registration Authority & \hspace{3pt} under development& & \\
 & - IBFT consensus & & & \\
\hline
Ethereum 2.0 & -  main chain (staking) & - cross-shard communication & - PoC in development. & Ethereum \\
Sharding \cite{EthereumFoundation2018} & - beacon chain (random numbers) & \hspace{3pt} via beacon chain, with & - production implementation & \\
 & - shard chains (data) & \hspace{3pt} detailed specification still & \hspace{3pt} after Ethereum 1.x & \\
 & -  PoS consensus (BFT) & \hspace{3pt} under development & &\\
 & - participant roles: proposers, & & & \\
 & \hspace{3pt} attesters, validator committees & & & \\
\hline
Metronome \cite{Metronome2018}& - Combined PoS \& PoW  & - interoperability via & partially implemented, some & Ethereum,  \\
 & \hspace{3pt} consensus & \hspace{3pt} export receipt \&  & functionality still under & potentially \\
 & - unclear whether public & \hspace{3pt}  import redemption & development & platform \\
 & \hspace{3pt} or private & - cross-chain token (MET) & & agnostic \\
\hline
NEC Laboratories \cite{Li2017b}&- Run BFT consensus protocols  & - direct asset transfer & - PoC (pseudo code provided)  & Hyperledger \\
 & \hspace{3pt} (unspecified) in parallel & \hspace{3pt} via sidechains (using TLS)  & - PoC integrated with HF & Fabric (HF)\\
 & - `Hands-off' regulator &  & - website markets `satellite chains' & \\
 & \hspace{3pt} overseeing network  & & \hspace{3pt} as a product & \\
 & - private and permissioned & & & \\
 \hline
Plasma MVP & - UXTO-based chain & - smart contract for deposits & - in production: Jan 2018 & Ethereum \\
(Minimal Viable Plasma) & - PoA consensus & \hspace{3pt} and withdrawals  & - implementation by OmiseGO & \\
 \cite{Poon2017,Buterin2018,LearnPlasma,Konstantopoulos2018,Knott2018} & - majority of work `off-chain' & & \hspace{3pt} for a decentralised exchange & \\
 & - tree hierarchy \& MapReduce & & &  \\
 \hline Plasma Cash & - non-fungible tokens (ERC721)& - propose PlasmaChain as& - in production on Mar 2018 & Ethereum \\
 \cite{Buterin2018a, Duffy2018} & \hspace{3pt} ETH \& ERC20 tokens& \hspace{3pt} parent chain for cross-  & - three sidechains in production: & \\
 & - sparse Merkle tree & \hspace{3pt} chain transactions, without & \hspace{3pt} PlasmaChain (aka ZombieChain), & \\
 & & \hspace{3pt}  pinning back to MainNet &\hspace{3pt} GameChain \& SocialChain & \\
 \hline PlasmaXT & - reduced storage requirements& & - in production: May 2018 & Ethereum \\
 \cite{Zhang2018, Fichter2018} & - aggregated signatures & & & \\
 & - checkpointing & & & \\
 \hline PlasmaDebit & - exchange of divisible amounts & & - in production: Jun 2018 & Ethereum \\
 \cite{LearnPlasma} & & & & \\
 \hline PlasmaMoreVP & - confirmation signatures  & & - in production: Jun 2018 & Ethereum \\
 (More Viable Plasma) & \hspace{3pt} no longer required & & & \\
 \cite{LearnPlasma} & - reorganised exit priority &  &  & \\
 \hline
Polkadot & - participant roles: nominators, & - interoperability via relay chain& PoC-1: May 2018 & Ethereum  \\
\cite{Wood2017,Wood2017a,Web3Foundation2017} & \hspace{3pt} validators (rewarding \& &  - inter-chain message & PoC-2: Krumme Lanke (Jul 2018) & used for PoC,\\
&  \hspace{3pt} slashing), fishermen, collators  & \hspace{3pt} passing (ICMP) still & PoC-3: Alexander (Jan 2019) & but plan to be \\
& - topology: parachains, & \hspace{3pt} in development & Substrate built using Rust \& LLVM & platform\\
& \hspace{3pt} relay chains, bridges & - ICMP PoC implementation & & agnostic\\
& - PoS consensus &\hspace{3pt} scheduled for April 2019 & & \\
 \hline
TAST \cite{Borkowski2018d, Borkowski2018, Borkowski2018c} & - cross-chain token (PAN)  & - interoperability via PAN tokens & PoC for cross-chain asset transfer & Ethereum \\
 & - transferable \& blockchain  & - transfer PAN between chains &  using claim-first transaction with & used for PoC, \\
 & \hspace{3pt} independent token& - atomic token transfers & cryptographic proof of intent& but plan to be \\
 & - PAN used to trade assets & - claim first transactions & & platform \\
 & \hspace{3pt} between native currencies & \hspace{3pt} with deterministic witness& & agnostic\\
\hline
\end{tabular}
\label{tab1}
\end{center}
\end{table*}

\section{Conclusion}
As is evident from Table \ref{tab:summary}: \textit{\nameref{tab:summary}} on page \pageref{tab:summary}, there are various `flavours' of blockchains with a steady stream still entering the market. Despite individual variations in setting up and running sidechains, the overarching concepts are very similar. The sheer number of public and permissioned blockchains is driving a growing desire and urgency to be able to transact and communicate amongst these disparate blockchains. However, researchers and developers recognise that blockchain interoperability is very challenging and complex. Several solutions have been proposed, some of which are summarised in the table, and each has its own unique set of problems to solve, but in essence there are only a few high level approaches to interoperability. A common strategy is facilitating message passing via a management chain, e.g. Ethereum 2.0 beacon chain, and Polkadot's relay chain, others use a decentralised exchange (DEX), e.g. for Clearmatics atomic swaps, and blockchains such as Metronome and the TAST project use a cross-chain cryptocurrency for atomic asset transfers. Another strategy is direct communication via hardware connections using TLS, as implemented by NEC, or via smart contracts using atomic function calls, as proposed by Ethereum private sidechains. \\
\section*{Acknowledgment}
This research has been undertaken while the authors have been employed by ConsenSys. We would like to thank Ben Edgington for his careful review and helpful comments.
\bibliographystyle{IEEEtran}
\bibliography{IEEEabrv,sidechains}

\end{document}